# Tuning topologically nontrivial states in the BHT-Ni metal organic framework

Nafiseh Falsafi,[1] Saeed H. Abedinpour,[2] Fariba Nazari*,[1,3] Francesc Illas*[4]

[1]Department of Chemistry, Institute for Advanced Studies in Basic Sciences, Zanjan 45137-66731, Iran

[2]Department of Physics, Institute for Advanced Studies in Basic Sciences, Zanjan 45137-66731, Iran

[3]Center of Climate Change and Global Warming, Institute for Advanced Studies in Basic Sciences, Zanjan 45137-66731, Iran

[4]Departament de Ciència de Materials i Química Física & Institut de Química Teòrica i Computacional (IQTCUB), Universitat de Barcelona,C/Martí i Franquès 1, 08028 Barcelona, Spain

**Abstract**

Using first principles calculations, we have demonstrated the creation of multiple quantum states, in the experimentally accessible metal organic framework BHT-Ni. Specifically, quantum spin Hall and quantum anomalous Hall states are induced by two and four electron doping, respectively. The geometrical symmetry breaking, is also investigated. For a low electron doping concentration of two electrons per unit cell, the Fermi energy shifts to a nontrivial band gap, between Dirac bands and a quantized spin Hall conductivity is predicted. Subsequently in a high electron doping concentration, Anomalous Hall conductivity with a quantized value was observed. In addition, for centrosymmetric (trans-like) and non-centrosymmetric (cis-like) structures, we found that the trans-like structure preserves quantum spin Hall and quantized spin Hall conductivity. In contrast, in the cis-like structure, space inversion symmetry breaking leads to the appearance of valley Hall effect and the disappearance of spin Hall conductivity.

*Corresponding authors: nazari@iasbs.ac.ir, francesc.illas@ub.edu

## 1. Introduction

Quantum materials constitute a suitable and information-rich playground for the simultaneous study of topology and symmetry and their mutual influence, owing to the profound interplay between them. Diverse rich spectrum of electronic and topological states and definite symmetry arise from the strong interconnection of lattice-, charge-, spin-, orbital- and valley degrees of freedom (DOFs) [1-3]. The complex interactions among these DOFs create a dynamic interconnection between them offering insights into their potential applications [4]. Among these interactions, spin-orbit coupling (SOC) reveals exciting possibilities for spin devices which operate without magnetic fields [5,6]. Indeed, in some systems with strong spin-orbit coupling, topologically-protected helical edge or surface states exist [7]. Along with the bulk electronic energy bands, these states define various quantum or topological phases of matter [8].

While significant efforts have been made to discover new topological materials with exotic phases, the practical application of these properties in devices remains limited. This limitation arises from the slow progress in experimentally synthesizing structures capable of controlling topological phase conversions, which often require extreme tuning conditions or complex multi-layered topological configurations [9]. In this context, metal-organic frameworks (MOFs) represent a promising class of materials that could significantly broaden the horizons of quantum material design. In last decade significant breakthroughs in manufacturing atomically layered two-dimensional (2D) MOFs with Kagome lattice structures have been reported [10-12]. Some of these structures have been theoretically predicted and experimentally confirmed [13] as materials with topological properties [14-19]. For instance, the electronic structure of a Kagome lattice naturally features Dirac fermions, which carry information about topology, flat bands (FBs) that are associated to correlated phenomena such as magnetism, and van Hove singularities which can induce instabilities towards long-range many body orders [20].

Although various lattices demonstrate the presence of FBs, the Kagome lattice stands out due to its practicality from a synthetic perspective as does not require tuning hopping parameters to host a FB [21]. Note, however, that FB can be topologically trivial (FB) or topologically nontrivial (TFB) [22-24]. Also, strong SOC may induce a nontrivial gap opening, which leads to an isolated TFB [25], with a nonzero (spin) Chern invariant [26]. The emergence of TFB adds further interest to the topological properties' investigation [27].

On the other way, it has been shown that topological phases such as two-dimensional topological insulators (2D TIs) which are named quantum spin Hall (QSH), and quantum anomalous Hall (QAH) can be controlled through an external electric field. It enables precise manipulation of the lattice structure, charge distribution, orbital configurations and spin orientations which opens new pathways for engineering electronic and topological properties [28-30]. Owing to absence of the Fermi level, exactly within the nontrivial band gap, the precise control of the Fermi level is most often crucial for identifying topologically nontrivial states in organic topological insulators (OTIs) [31]. As an example, the 2D planar MOF, $Ni_3C_{12}S_{12}$, also referred to as (BHT-Ni) [10], is the first experimentally realized 2D OTI, which its Fermi level is located within the trivial band gap. Here, one requires a level of electron doping equivalent to two electrons (per unit cell) ~$5\times10^{13}$ cm$^{-2}$ [14,32] or four electrons (per unit cell) which is equal to ~$2\times10^{14}$ cm$^{-2}$ to shift the Fermi level to topologically nontrivial gap [14]. Doping with electrons or holes has also been proposed for other topological phases [33,34]. It is essential to highlight that doping primarily adjusts the Fermi level position only, while the OTI electronic band topology remains unaffected.

Additionally, the manipulation of the Fermi level has been shown to effectively enhance the spin Hall conductivity (SHC) [35]. A prominent method to achieve SHC is through the spin Hall effect (SHE), which converts a charge current into a spin current [36]. The SHC can be categorized into intrinsic, directly derived from the relativistic band structure [37] or as extrinsic [36]. Due to the spin-momentum-locked surface states, topological insulators are considered as ideal materials for generating a pure spin current with a large spin Hall angle (SHA). SHA is the figure of merit of charge-to-spin current interconversion of spin-orbit torque (SOT) [38] devices. In this respect, finding new materials with large SHA is decisive [39].

One of the most fascinating topological phases is the anomalous Hall effect (AHE), which usually accompanies time reversal symmetry (TRS) breaking. An insulator featuring the quantum mechanical version of anomalous Hall effect (QAH), exhibits a topologically nontrivial electronic structure with a finite Chern number recognized by the bulk energy gap and gapless chiral edge states. This unique property results in quantized Hall conductance even in the absence of an external magnetic field [40,41]. Note that QAH, can also be realized in magnetically doped TIs or intrinsic magnetic topological insulators [42-44]. However, magnetic doping degrades sample quality and limits the QAH state's critical temperature [45]. Thus, it will be important to explore new methods for realizing this phase. As mentioned above for SHC, anomalous Hall conductivity (AHC) arising from AHE can be adjusted by changing the chemical potential through electron or hole doping. Additionally, the transition from QSH to QAH states induced by electron doping has been observed in some OTIs [16]. High-throughput calculations of AHC in 2871 ferromagnetic materials revealed that those with large AHC (greater than ~1000 S/cm) are rare, with most metals showing AHC between 10 and 1000 S/cm [46]. Kagome magnets are known for their large AHC [47-51]. Beside the role of electron and spin DOFs, in electronics and spintronics, respectively, the valley DOF can lead to the emergence of a new field which is known as valleytronics. It exploits the valley index of electrons to store and manipulate information. Structures with two-dimensional honeycomb lattice have valley DOF, in addition to the charge and spin. But in the presence of space inversion symmetry (SIS), the electric and optical control of valley DOF is difficult [52]. In contrast breaking the SIS in two-dimensional honeycomb structures, opens a finite gap in the Dirac cones, at high symmetry K and K′ points of the reciprocal space, with the opposite sign of local Berry curvature at these two points [53]. Due to the broken SIS, electrons in two valleys experience opposite effective magnetic fields that are proportional to to the Berry curvature. These fields cause electrons from different valleys encounter opposite Lorentz-like forces and move in opposite directions, perpendicular to an applied current. In this case there is valley polarized carriers which is similar to the spin Hall effect (SHE), where spin-polarized electrons move in opposite directions. This phenomenon called the valley Hall effect (VHE) [52].

Based on the previous summary of the existing literature, we present a comprehensive analyses of electronic, transport and topological properties of the electron-doped organometallic framework (BHT-Ni)$_p$. We show that at low electron doping concentration ($1.07 \times 10^{14}$ cm$^{-2}$), the trivial insulator (BHT-Ni)$_p$ transitions into a QSH insulator which is characterized by a nonzero $Z_2$ topological invariant. While at a higher electron doping level ($2.15 \times 10^{14}$ cm$^{-2}$), it transitions into a QAH insulator with a finite Chern number of $C = 1$, due to TRS breaking and spontaneous spin polarization. Altogether, we have analyzed, at experimentally assessable doping levels, multiple transitions from a trivial insulator [54] to a QSH insulator, and then to a QAH insulator, all achieved within a single material, confirming the prediction of Zhang et al. [16]. Furthermore, we have identified and analyzed salient characteristics of topological transport like SHC and AHC of pristine and electron doped BHT-Ni. Finally, the impact of geometrical symmetry breaking on the electronic and topological behavior are investigated in centrosymmetric (trans-like) and non-centrosymmetric (cis-like) structures. These structures are created by substitution of sulfur ligands with iso-valence selenium ligands. By symmetry reduction trans-like structure preserve QSH and quantized SHC. While in cis-like structure SIS breaking leads to appearance of VHE and SHC disappears.

## 2. Computational details and material modeling

The calculation of the electronic structure and topological properties of the BHT-Ni systems examined in this study, was performed within the framework of density functional theory [55] using the Quantum ESPRESSO package [56]. A plane-wave basis set, augmented with ultrasoft and projector augmented wave pseudopotentials [57] was employed in the calculations. The exchange and correlation functional were treated using the generalized gradient approximation of Perdew-Burke-Ernzerhof [58], which is widely recognized for its accuracy in describing the electronic structure of materials. The kinetic energy cutoff for the plane-wave basis was chosen to be 80 Ry. A ($3 \times 3 \times 1$) *k*-point mesh was used for the BZ sampling. The SOC effect was taken into account for the calculation of electronic structures and topological properties.

Models for pristine, low and high electron doping concentrations, cis- and trans-like configurations were used, which are denoted as $(BHT-Ni)_p$, $(BHT-Ni)_l$, and $(BHT-Ni)_h$, cis-$(BHT-Ni)_p$ and trans-$(BHT-Ni)_p$, respectively. In the structural optimization, the internal atomic positions were relaxed until the forces on all the atoms were smaller than 0.002 Ry/Bohr. In our first-principle calculations, the doping effect is simulated by adding electrons to the lattice, and meanwhile adding a homogeneous background charge of opposite sign to maintain the system charge neutrality.

To investigate the stability of the (BHT-Ni)$_p$ monolayer structures, we calculated the cohesive energy (E$_C$) as eq 1:

$$E_C = \frac{\left[E_{Ni_3C_{12}X_aY_b} - (N_{Ni}E_{Ni} + N_C E_C + N_X E_X + N_Y E_Y)\right]}{N} \qquad (1)$$

where $E_{Ni_3C_{12}X_aY_b}$ is the total energy of the monolayer of interest, $E_{Ni}$, $E_C$, $E_X$ and $E_Y$ are the total energies of the free Ni, C, X and Y atoms, respectively. In (BHT-Ni)$_p$ structure, X = Y = S while in cis- and trans-(BHT-Ni)$_p$, X = S and Y = Se. $N_{Ni}$, $N_C$, and $N_X$ = a and $N_Y$ = b are the number of Ni, C, X and Y atoms, respectively. N stands for the total number of atoms in the monolayer structures.

The Bloch states arising from the periodic calculations, are next expanded in the Wannier basis by projecting into p$_z$ orbitals of C and S together with d$_{xz}$ and d$_{yz}$ orbitals of Ni, which comprise the Kagome bands in the vicinity of the Fermi level. The initial projected Wannier functions are then optimized to obtain a maximally localized Wannier functions (MLWF) via WANNIER90 code [59], then the MLWF are used to derive the tight-binding Hamiltonian in these localized bases [60]. Utilizing the post processing module (BERRY) of WANNIER90 code developed by Qiao et al. [61] the SHC and SHA are calculated. The SHC that appeared in the Kubo formula approach (see section 3) is achieved by integrating the Berry-like curvature over the BZ [36,61]. For the Wannier interpolation, we chose 500 × 500 × 500 $k$-mesh grid for SHC calculations. An adaptive refinement $k$-mesh of 4 × 4 × 4 was used in the integral calculations. The method of adaptive $k$-mesh refinement can be effective for an efficient convergence of the SHC calculation [61]. The semi-infinite edge states were calculated using real space Hamiltonian in the basis of MLWFs and iterative Green's function approach implemented in WANNIERTOOLS [62].

## 3. Main concepts related to topological properties

By examining how band structures evolve under spin-orbit interaction, many topologically interesting phases of matter have been predicted, like QSH and QAH phases, among others. Two-dimensional TI or QSH insulator is an exotic state of quantum matter, distinct from the ordinary insulator phase by which electrons cannot conduct in the bulk of the material. While its edges support conducting electronic states (helical edge states) that are protected by TRS. The QAH insulator is another exotic phase of matter that is driven by TRS breaking due to the magnetic polarization [63]. Gapless chiral edge states in QAH insulator emerge due to this symmetry breaking. QAH and QSH states are characterized by quantized anomalous Hall conductivity (AHC) and spin Hall conductivity (SHC). The AHC and SHC can be separated into intrinsic parts directly derived from the relativistic band structure [37] and the extrinsic mechanisms where electrons acquire transverse velocity through skew or side jump scatterings [36]. The intrinsic AHC and SHC can be calculated by integrating respectively the ordinary Berry curvature (BC) and spin Berry curvature (SBC) of the occupied bands over the BZ [61]. The concept of Berry curvature is related to Berry phase that is a phase angle (i.e., running between 0 and $2\pi$) and describes the global phase acquired by the complex state vector as it is carried around a path in its vector space [64].

In a pretty compact way, the charge and spin conductivities that relate the applied electric field to the induced charge or spin currents could be written as $J_\alpha^\gamma = \sigma_{\alpha\beta}^\gamma E_\beta$ where the lower indices $\alpha$ and $\beta$ refer to the spatial directions of the induced current and applied field, respectively. The upper index, $\gamma$, refers to the charge degree of freedom ($\gamma = c$) for the charge conductivity or to the spin-polarization direction ($\gamma = x, y, z$) for the spin conductivity. Different components of the conductivity tensor $\sigma_{\alpha\beta}^\gamma$ could be calculated using the Kubo formula. In the modern language, the Kubo formula for the transverse conductivities in the direct current (dc) limit are conveniently expressed in terms of the BC and SBC [65], as

$$\sigma_{\alpha\beta}^{\gamma} = -\frac{e^2}{\hbar} \int_{BZ} \frac{d\mathbf{k}}{(2\pi)^d} \Omega_{\alpha\beta}^{\gamma}(\mathbf{k}). \qquad (2)$$

Here, $d$ is the dimension of the system and the total $\mathbf{k}$-resolved (ordinary and spin) Berry curvatures $\Omega_{\alpha\beta}^{\gamma}(\mathbf{k})$, are

$$\Omega_{\alpha\beta}^{\gamma}(\mathbf{k}) = \sum_n f(E)\, \Omega_{n,\alpha\beta}^{\gamma}(\mathbf{k}), \qquad (3)$$

where the sum is over bands $n$, $f(E) = 1/[e^{(E-\mu)/(k_B T)} + 1]$ is the Fermi-Dirac distribution function with $\mu$, $k_B$, and T, the chemical potential, Boltzmann constant, and absolute temperature, respectively. The Fermi-Dirac factor restricts the sum to occupied bands. The band-projected Berry curvature-like term are defined as [61]

$$\Omega_{n,\alpha\beta}^{\gamma}(\mathbf{k}) = -2\hbar^2 Im \sum_{m\neq n} \frac{\langle n\mathbf{k}|\hat{\mathcal{J}}_\alpha^\gamma|m\mathbf{k}\rangle\langle m\mathbf{k}|\hat{v}_\beta|n\mathbf{k}\rangle}{(\epsilon_{n\mathbf{k}}-\epsilon_{m\mathbf{k}})^2}, \qquad (4)$$

where $\epsilon_{n\mathbf{k}}$ and $\epsilon_{m\mathbf{k}}$ are the eigenvalues corresponding to the Bloch eigenstates $|n\mathbf{k}\rangle$ and $|m\mathbf{k}\rangle$. The spin-velocity operator is defined as $\hat{\mathcal{J}}_\alpha^\gamma = \{\sigma_\gamma, \hat{v}_\alpha\}/2$ where $\hat{v}_i = \hbar^{-1}\partial\hat{H}/\partial k_i$ with i = $\alpha$, $\beta$ is the velocity operator ($\hat{H}$ is the Hamiltonian), and $\sigma_\gamma$ with $\gamma$ = x, y, z are the Pauli matrices. Taking $\sigma_c$, a (2 × 2) identity matrix, we find $\hat{\mathcal{J}}_\alpha^c = \hat{v}_\alpha$ and can define ordinary and spin Berry curvatures in the same manner. Although the third-rank tensor $\sigma_{\alpha\beta}^\gamma$ can, in principle, have several elements, symmetry constraints fix the number of its independent elements [66,67]. Recently, symmetry is utilized to determine the nonzero components of the SHC tensor and simplify the calculations in Weyl semimetals [35] and topological insulators [68]. According to the symmetry analysis the allowed independent SHC components for all 230 space groups are tabulated in Table 1 [66]. On the other hand, in two spatial dimensions, which is the focus of our study here, charge and spin Hall conductivities each can have at most a single independent non-vanishing element, i.e., $\sigma_{xy}^c = -\sigma_{yx}^c$ and $\sigma_{xy}^z = -\sigma_{yx}^z$. Note that, the unit of Berry curvature $\Omega_{\alpha\beta}^\gamma(\mathbf{k})$ is length², and therefore in the way we have defined $\sigma_{\alpha\beta}^\gamma$ through eq 2, the units of both AHC and SHC reads $e^2/\hbar \, length^{(2-d)}$ ($d$ = 2, 3 is dimension of the system). While this is the standard unit for charge conductivities, to convert the SHC into its conventional units $e \, length^{(d-2)}$, it should be scaled by $-\hbar/(2e)$ (note that $e^2/\hbar \simeq 2.434 \times 10^{-4}$ S). Ab initio calculation of topological conductivities based on the Wannier method first introduced by Wang et al. [69] for the AHC. Subsequently, this method was expanded to examine the SHC as well. To evaluate the SHA here, we have adapted a two-dimensional version of the $\theta_{SH}$ [61],

$$\theta_{SH} = \frac{2e}{\hbar} \left(\frac{\sigma_{xy}^z}{\sigma_{yy}}\right) \tag{5}$$

the longitudinal electrical conductivity $\sigma_{yy}$ should be also calculated. To do that, we use the semiclassical Boltzmann transport equation [70]

$$\sigma_{\alpha\beta}(\mu, T) = e^2 \int_{-\infty}^{\infty} dE \left(-\frac{\partial f(E)}{\partial E}\right) \Sigma_{\alpha\beta}(E). \tag{6}$$

Here,

$$\Sigma_{\alpha\beta}(E) = \int \frac{d\mathbf{k}}{(2\pi)^d} v_\alpha(n, \mathbf{k}) v_\beta(n, \mathbf{k}) \tau_{n,\mathbf{k}} \delta(E - E_{n,\mathbf{k}}) \tag{7}$$

is the transport distribution function and $\langle n\mathbf{k}|v_i|n\mathbf{k}\rangle$, where i = $\alpha$, $\beta$ is the band velocity. Moreover, $\tau_{n,\mathbf{k}}$ is the relaxation time that generally depends on the band index and the wave vector. However, we resort to the constant relaxation time approximation i.e., $\tau_{n,\mathbf{k}} = \tau$, throughout this work.

## 4. Crystal structure and main symmetry features

A benzene hexathiol (BHT) molecule, contains a benzene core and three chelating dithiolene ligands. The symmetrical structure of the core and the practicality of covalent metal dithiolene bonding makes it an excellent molecular building block in the construction of porous polymer MOFs, such as the BHT-Ni that was achieved through a coordination reaction between benzene hexathiol (BHT) and nickel (II) acetate (Ni-(OAc)$_2$) [10]. In BHT-Ni which we named (BHT-Ni)$_p$, every BHT unit bonds with three Ni atoms to build a structurally perfect Kagome lattice pattern that depicted with black dashed lines in Figure 1a.

Benzene Hexathiol-Ni (BHT-Ni)$_p$ that also named Ni-bis-dithiolene Ni$_3$(C$_6$S$_6$)$_2$ has a planar configuration and sixfold symmetry as a consequence of structural relaxation. In addition, first principle simulations demonstrated both its thermal stability [71] and the dynamic stability [72]. The metal sites in these M$_3$L$_2$-type MOFs have a local structure of four neighboring atoms in a planar configuration regardless of the type of ligand molecules. For example, the Ni atom in (BHT-Ni)$_p$ (Figure 1a) is surrounded by four sulfur atoms [73]. The (BHT-Ni)$_p$ material classify as a hexagonal crystal with space inversion symmetry (SIS), the symmetry of the crystal corresponds to the P6/mmm (No. 191) space group, which is homomorphic to the point group D$_{6h}$. Its crystal structure generates by a threefold rotation symmetry (C$_{3z}$), the vertical mirror symmetry —because of the threefold rotation symmetry, there are three equivalent vertical mirror symmetries $M_{vi}$ (i=1,2,3) as shown in Figure 1b—, the horizontal mirror symmetry ($M_h$) and the SIS.

The unit cell of (BHT-Ni)$_p$ contains 3 Ni atoms, 12 C atoms and 12 S atoms that depicted with red lines in Figure 1a. This 2D d$^8$ planar transition metal containing (BHT-Ni)$_p$, is known to display a high electrical conductivity (EC) up to 160 S/cm [10,74] and realize efficient π−d conjugation and delocalized electrons [54]. The optimized lattice constant of the (BHT-Ni)$_p$ monolayer is $a = b = 14.64$ Å, which is in good agreement with recent experimental measurements [10] and previous first principle DFT results [14,71]. The calculated band structure without considering SOC and the corresponding projected density of states (PDOS) are displayed in Figure 1c. These clearly illustrate the semiconductor character and the contribution of constituent atoms in electronic bands of $(BHT-Ni)_p$. The agreement between the results from present DFT calculations for the crystal structure and the previously reported data [14,32,71,75,76] confirms the suitability of the chosen computational methods and parameters to proceed the investigation.

## 5. Electronic and topological properties in the pristine and low electron doping regime

Before proceeding with the results, it is important to highlight that the studied structure $(BHT-Ni)_p$ has been confirmed to be thermodynamically stable. The cohesive energy values (eq 1), which underscore this stability, are reported in Table 2 and have been methodically compared with available data to ensure the accuracy and reliability of our results. Additionally, the stability of the studied structures has been compared with that of graphene, providing a benchmark to further validate our findings.

Since all the calculations for the structures mentioned in the computational section, require information about the pristine structure, we will first present the calculations and details obtained for the $(BHT-Ni)_p$. Then, we will present the results and analyses for the charged counterparts. By means of what was already mentioned, $(BHT-Ni)_p$ is a semiconductor with a narrow indirect energy gap of 0.122 eV. Figure 1c demonstrates that Kagome bands (*cf.* red bands in Figure 1c left) comprising one FB above two Dirac bands, that are located above the Fermi level ($E_F$), within $E_F < E < E_F + 0.8$ eV [32,75,76]. These are graphene-like energy bands, with a Dirac cone at the K point, degenerated with a FB at the Γ point. Note, however, that these degeneracies will be removed by the SOC and TFB will be appear (see Figure S2a of the SI). In fact, once accounting for SOC in (BHT-Ni)$_p$ we found a nontrivial global energy gap of $\Delta_3 = 5.4$ meV and a local energy gap of $\Delta_2 = 17$ meV at the Γ point and $\Delta_1 = 14$ meV Dirac energy gap at the K point which are all in good agreement with recent first principle results for global gap at the range of $\Delta_3 = $ 4-5.8 meV [14,75] and $\Delta_2 = 17$ meV [32] for local gap at the Γ point and $\Delta_1 = $ 13.6-14 meV for Dirac gap [14,32,75]. These gap openings are mainly due to intrinsic SOC within the d-orbitals of Ni atoms, herein, Rashba SOC effect can be excluded due to the inherent lattice inversion symmetry. These SOC gaps are much larger than the gap opened in graphene [77]. The enhanced SOC interactions here are attributed to the hybridization of p$_z$ orbitals of light atoms with d orbitals of Ni atoms. The SOC gaps of OTIs are typically in this range (2.3–255 meV) [33, 78]. The paired Ni-3d electrons, do not break the TRS of the original (BHT-Ni)$_p$ structure and consequently, a nonmagnetic ground state is predicted [73], where the SOC promotes the emergence of the QSH phase as will be shown below. A crucial property of TRS, when applied to electrons with significant spin properties, is expressed by the Kramer's theorem, which states that all energy eigenvalues are at least two-fold degenerate [64]. This is what one would expect from spin degeneracy in a nonmagnetic system in the absence of SOC; the Kramer's theorem states that the bands remain doubly degenerate even in the presence of SOC if SIS is also present. We calculated the projected band structure of (BHT-Ni)$_p$ that is reported in Figure S1. This clearly shows that the Kagome bands come mainly from the Ni (d$_{xz}$, d$_{yz}$), C and S p$_z$ orbitals.

It is known that having the Fermi level (E$_F$) precisely situated within the nontrivial gap of the material, is beneficial for experimental measurements of electronic transport properties. Therefore, we tried to raise E$_F$ to the Kagome band region by electron doping. Based on first principle calculations at a doping concentration of $1.07 \times 10^{14}$ cm$^{-2}$, corresponding to two additional electrons per unit cell, E$_F$ shifts exactly to the Dirac point, without changing Kagome bands, as evident by comparison of Figure 2a and 2b. The projected band structure of (BHT-Ni)$_l$, in Figure 2c illustrates that electron doping does not change the nature of Kagome bands. By electron doping, lattice constant slightly increased by about 0.9% but both symmetries, TRS (E ($k$, ↑) = E (-$k$, ↓)) and SIS (E ($k$, ↑) = E (-$k$, ↑)) are effectively retained. This leads to energy bands retaining their spin degeneracy, namely the Kramer's degeneracy [79]. It is worth noting that this doping regime is experimentally achievable [80]. The DFT calculations with including SOC, show that the degeneracies of the two Dirac bands at the K point, and of the upper Dirac band and TFB at the Γ point are lifted, leading to energy gaps of $\Delta_3$ = 8.5 meV as global gap and $\Delta_2$ = 18.6 meV as local gap at the Γ point and $\Delta_1$ = 15.4 meV at K point as Dirac gap, (see Figure 2d and S2b).

To investigate the topological properties, we calculated the edge states, $Z_2$ invariant and the Berry curvature of the (BHT-Ni)$_p$ and (BHT-Ni)$_l$ lattice by employing the MLWF-based Hamiltonian obtained from WANNIER90 [59] and postprocessing WANNIERTOOLS package [62]. Based on orbital resolved band structures of Figure S1 and Figure 2c, we constructed MLWFs using the p$_z$ orbitals of S and C atoms together with d$_{xz}$, d$_{yz}$ orbitals of Ni atoms as the initial guess for constructing tight binding Hamiltonian. This is a common approach used to study topological properties [75,81-83]. Due to its trivial gap in the case of the (BHT-Ni)$_p$, no edge state was observed at the Fermi level, and topological $Z_2$ invariant was zero. But the (BHT-Ni)$_l$ with non-trivial gap at the Fermi level, is expected to be receptive to the nontrivial topological phase. Figure 2e shows the fitted band structure of (BHT-Ni)$_l$ with Wannier interpolation in the energy window of (E$_F$ − 0.5, E$_F$ + 0.4) eV, that accurately matched the DFT bands. The edge states for the semi-infinite lattice in Figure 2f were calculated using real space Hamiltonian in the basis of MLWFs and iterative Green's function approach [62]. On each side of the sample a helical edge state appears, with both spin channel on right (Figure 2f, right) and left (Figure 2f, left) sides which connecting valence bulk band with conduction bulk bands. Figure 2f is the hallmark of QSH insulators, that is the bulk states are connected by topologically nontrivial edge states. $Z_2 = 1$ is the direct output of our calculations using the WANNIERTOOLS package. So, (BHT-Ni)$_l$ is a QSH insulator (two dimensional TI) or due to its organic structure, is a two dimensional OTI. The possibility of using various metal atoms and molecular ligands makes organic topological materials highly tunable, which is their advantage [14].

At this point it is worth exploring the relationship between SHC and crystal symmetry. The SHC describes the generation of spin currents perpendicular to an applied electric field in materials with spin-orbit coupling. The relationship between SHC and crystal symmetry is crucial, as symmetry properties strongly influence the spin-orbit interactions responsible for SHC. Symmetry dictates the allowed forms of spin-orbit coupling in a crystal, which in turn affects the magnitude and even the directionality of SHC. For instance, certain crystal symmetries may enhance or suppress spin-orbit interactions, thereby influencing the efficiency of spin current generation [84]. For this purpose, by employing MLWF, at first we calculated the SHC of the (BHT-Ni)$_p$ and later for (BHT-Ni)$_l$. The P6/mmm crystal structure of (BHT-Ni)$_p$ corresponds to the nonmagnetic Laue group 6/mmm; it has three independent elements of SHC at three dimension ($\sigma_{xy}^z$, $\sigma_{zx}^y$, $\sigma_{xz}^y$) [66] and only one at two dimension ($\sigma_{xy}^z$). This is due to the 2D structure of (BHT-Ni)$_p$ causing the charge and spin currents to be confined within the *xy* plane [84].

It is obvious that, in experimental setups, the best method to validate the QSH is to measure the SHC as a function of gate voltage. This allows for tuning the chemical potential using spin-dependent transport measurements. In this section, we will investigate the impact of Fermi level variations by electron doping on SHC. First principle calculations using density functional theory are performed to correlate the band structure with SHC. Figure 3a illustrates the band structure of (BHT-Ni)$_p$ with including the SOC and Figure 3b the corresponding SHC, $\sigma_{xy}^z$ and its quantization at two nontrivial band gaps (Γ and K point) that are not at the Fermi level. Note that both the band structure and SHC are obtained in the Wannier basis. The value of SHC ($\sigma_{xy}^z$) of (BHT-Ni)$_p$ at the Fermi level is -0.093 ($\hbar$/e) (S/cm), actually it is near zero due to the trivial gap combined with both the SIS and TRS of the system. Upon up-shifting E$_F$ by assuming electron doping, the absolute value of the SHC starts to increase. When the E$_F$ reaches the top of the lower Dirac band, the value of the SHC reaches its maximum value and keeps it until E$_F$ is within the Dirac gap. Therefore, a SHC plateau achieves, which can be described by spin-polarized edge states in the QSH phase. When E$_F$ is placed near the bottom of the upper Dirac band, the SHC quickly reduces and then increases again when E$_F$ crosses the upper Dirac band. During this process, the SHC arrives to another maximum value again, when the E$_F$ is within the second nontrivial gap. However, now the width of the plateau is less, depending on the size of the nontrivial global band gap. Outside of these peaks, the SHC quickly reduces again, as E$_F$ moves through the gap $\Delta_3$. The SHC reaches zero when the three Kagome bands are completely filled. Figures 3a and 3b show that moving E$_F$ to the nontrivial gaps, can increase SHC and reach the maximum values within both SOC gaps. By adding two electrons per unit cell and raising the Fermi level (Figure 4a), the value of SHC reaches to the maximum value of the -325 ($\hbar$/e) (S/cm) at SOC gap in (BHT-Ni)$_l$ (see Figure 4b). This value is of the same order of magnitude as other topological insulators [68]. As expected, SHC in 2D topological insulators should be quantized. In section A of SI you can see the calculation of quantized (in the units of $\frac{e}{2\pi}$) value of SHC in (BHT-Ni)$_l$. This quantized value is observed at two nontrivial gaps.

A more comprehensive understanding of the origin of the finite SHC inside the nontrivial gap can be obtained by analyzing the contribution of the bands in the E$_F$ vicinity. Figure 5a and 5b illustrates the band projected SBC, *i.e.* $\Omega^z_{n,xy}(\boldsymbol{k})$ at eq 4, for energy bands close to the SOC gap in (BHT-Ni)$_p$ and (BHT-Ni)$_l$, respectively. The color of the bands corresponds to the sign and magnitude of the SBC, *i.e.* sgn ($\Omega^z_{n,xy}(\boldsymbol{k})$) log $|\Omega^z_{n,xy}(\boldsymbol{k})|$. Figure 5b depicts a large contribution of conduction and valence bands to the SBC, especially in the vicinity of the Fermi energy. Furthermore, the sign of $\Omega^z_{n,xy}(\boldsymbol{k})$ changes rapidly as the energy approaches the SOC gap. The significant value of the SBC and its rapid sign change at E$_F$ suggest that the SHC is related to the topological order of the bands [68]. We also note that previous calculations on trivially gapped semiconductors [85] have revealed a nonzero residual SHC within the gap; however, its source does not appear to be topological. Hence, the bulk in TIs could produce a finite spin current even if the Fermi level is situated within the gap. A finite SBC and the sign flip at the Γ point in Figure 5a and 5b demonstrate its topological nature, which is confirmed by the quantized SHC with a small plateau. The small plateau is due to smaller SOC gap at the Γ point compared to the SOC gap at the K point, which is shown in Figure 4b around 0.2 eV.

We calculated the 2D distribution of $\boldsymbol{k}$-resolved SBC in eq 3 at E = E$_F$ for (BHT-Ni)$_p$ and (BHT-Ni)$_l$. The result of (BHT-Ni)$_l$ is shown in Figure 2g. But in the case of the (BHT-Ni)$_p$, the SBC was close to zero, so the SBC is highly sensitive to the position of the Fermi energy which is in agreement with recent reported data [35]. The red color in the $\boldsymbol{k}$-resolved SBC denotes positive values. These investigations explain the mechanism for SHC variation with Fermi energy, and outline the roadmap for adjusting the SHC.

We have also studied the SHA of (BHT-Ni)$_p$ and (BHT-Ni)$_l$ structures based on eq 5. To this end we calculated the electrical conductivity (EC) tensor components (in S/cm) which together to the corresponding SHC and SHA are reported in Table 3. The calculated SHA of 1.55 is comparable to that exhibited by other TIs 0.1 <θ$_{SH}$< 1.0 [68], the heavy metals such as platinum with high SHC (~2000 ($\hbar$/e)(S/cm)) and 0.056 < $\theta_{SH}$ < 0.16 [86], or 0.12 < $\theta_{SH}$ < 0.15 for tantalum [87]. Owing to the finite SHC and confined longitudinal charge conductivity, the SHA of TIs are comparable to that of heavy metals. Therefore, TIs are an ideal choice for energy-efficient charge-to-spin conversion, in spin-based devices. As mentioned previously, this implies that for the same spin current, TIs need a lower value of charge current in comparison with heavy metals, which have a fairly higher conductivity. In Table S1 of the SI, we summarized the SHA for some metals, semiconductors and TIs. SOT switching using the SHE in heavy metals and TIs holds significant potential for ultralow power magneto resistive random-access memory (MRAM). To be competitive with conventional SOT switching, a pure spin current source with a large SHA ($\theta_{SH}$ > 1) and high electrical conductivity ($\sigma$ > $10^5$ $\Omega^{-1}$ $m^{-1}$) is required [88]. Therefore, (BHT-Ni)$_l$ with $\theta_{SH}$ = 1.55 and electrical conductivity 3.4 × $10^5$ $\Omega^{-1}$ $m^{-1}$ could be a promising candidate in generated torque. In addition, the relatively high values of SHA for (BHT-Ni)$_l$ (1.55) indicate the high efficiency of charge-spin current conversion, which offers a great advantage for their applications in spin Hall devices.

## 6. Electronic and topological properties in the high electron doping regime

Wang et al. [14] and Zhang et al. [16] noted that, when the Fermi level is not within the SOC gaps, doping is an alternative method to achieve desired electronic and topological properties. Based on these observations, hereby we explore the resultant electronic, structural, and magnetic properties that emerge from high electron doping concentration in (BHT-Ni)$_p$. Hence, we further increased the electron doping concentration to 2.15×$10^{14}$ $cm^{-2}$ (four electrons in one unit cell), which is experimentally accessible as low electron doping concentration level. With this, we will provide a detailed understanding of how this doped system behave, potentially revealing novel properties, and eventually filling the knowledge gap left by Wang et al. [14].

Note that at this doping level, the lattice constant is extended only by less than 2% and, hence, the (BHT-Ni)$_h$ framework remains thermodynamically stable. From qualitative arguments, the Fermi level is anticipated to rise into the space between the TFB and the upper Dirac band, assuming that the Kagome bands remain unchanged. Nonetheless, our DFT analyses revealed that, at this specific doping level E (**k**, ↑) ≠ E (-**k**, ↓) and the TRS is broken, resulting in the emergence of spontaneous electron spin-polarization by magnetic moment of 2.0 $\mu_B$ per unit cell, as estimated from the spin density. Figure 6 illustrates the spin-polarized electron density, which is obtained from the difference in charge density between spin-up and spin-down channels δρ = ρ↑−ρ↓. Each sulfur atom exhibits a magnetic moment of 0.027 $\mu_B$, while each Ni atom shows a magnetic moment of 0.211 $\mu_B$, contributing approximately 16.20% and 31.65% to the overall magnetic moment, respectively.

Because of the overlap between the TFB and Dirac bands (see Figure 7a), electrons start to occupy the TFB as the Fermi level is elevated into this region. The instability caused by this, leads to spin polarization. Consequently, the spin degeneracy of the Kagome bands is lifted. In this scenario, the Kagome bands of one spin channel (spin-up) become fully occupied by electrons, while only one Dirac band of the other spin channel (spin-down) is filled. The Fermi level aligns precisely with the Weyl point of the spin-down channel, as illustrated in Figure 7b. While spin-polarized Dirac cones have been observed multiple times [89], to the best of our knowledge spin-polarized Dirac cones resulting from electron doping have only been reported by Zhang et al. in electron-doped HTT-Pt [16].

Considering SOC and spin-polarization, the present DFT computations predict the emergence of a band gap of 18 meV at the Fermi level, illustrated in Figure 7c. Using MLWFs based on the initial guess (obtained from projected band structure similar to Figure 2c) in the energy window ($E_F - 0.65$ eV, $E_F + 0.3$ eV), the Kagome bands can be accurately approximated using Wannier interpolation, as also depicted in Figure 7c. The heavy electron doped (BHT-Ni)$_h$ exhibits topological nontriviality, which can be demonstrated by calculating the Chern number ($C$). Figure 7d illustrates AHC as a function of the Fermi energy shift. It shows that the AHC forms plateaus in the units of $e^2/h$ at the Fermi level, signifying that the (BHT-Ni)$_h$, behaves as a Chern insulator with a nonzero Chern number ($C = 1$).

The distribution of the ordinary Berry curvature $\Omega_{xy}^c$ in 2D momentum space of the lower Dirac band (band I or valence band maximum in Figure 7c) is illustrated in Figure 7f. Because of the 2D structure of (BTH-Ni)$_h$, the calculated Berry curvature has only one nonzero component $\Omega_{xy}^c$. One can see that the nonzero Berry curvature is mainly localized around K and K′ points with the same sign, in agreement with the Chern number calculation. To provide additional validation of the topological nontriviality of (BHT-Ni)$_h$, we used the Green's function method to calculate the edge states of the spin-polarized (BHT-Ni)$_h$. The chiral edge states depicted in Figure 7e clearly show the characteristics of TRS breaking, as only a single spin channel links the bulk states. The calculated nonzero Chern number and topologically nontrivial edge states in the (BHT-Ni)$_h$ show great potential for realizing the QAH effect with a single spin-polarized edge channel. Although, QAH phase in TIs was frequently reported by doping with magnetic atoms [90] or through the introduction of proximity coupling with antiferromagnetic structures [91], here we accomplish this through electron doping. Similar to Zhang et al. which have suggested electron doping approach for inducing ferromagnetism in a TI and realization of Chern insulator [16]. The Chern number $C$ and AHC ($\sigma_{xy}^c$) are related as in eq 8 and can be obtained by integrating the Berry curvature over the first BZ.

$$C = \frac{1}{2\pi} \int_{BZ} d^2k\, \Omega_{xy}^c(\mathbf{k}) \qquad \text{and} \qquad \sigma_{xy}^c = -C \frac{e^2}{h} \tag{8}$$

The quantization of calculated AHC ($\sigma_{xy}^c = -323$ S/cm) in (BHT-Ni)$_h$, becomes evident when expressed in the units of ($\frac{e^2}{h}$),

$$\sigma_{xy}^c \text{ (S/cm)} = -323 \frac{S}{cm} \times 1.2 \times 10^{-7} cm \times \frac{e^2/h}{(3.874 \times 10^{-5})S} = -1 \frac{e^2}{h}$$

On each side of the sample a chiral edge channel appears, with opposite direction of propagation on right and left sides (Figure 7e) and the $C = 1$ value for the Chern number characterizes a quantized Hall conductivity, which confirm the anticipated topological nontriviality of the SOC gap. As noted, the magnitude of AHC at the Fermi level, reaches $\sigma_{xy}^c = -323$ S/cm or $\Omega^{-1}cm^{-1}$, which is comparable to the large AHC of the Kagome materials [41]. Hence, by controlling charge DOF the realization of different topological phases (QSH and QAH) at (BHT-Ni)$_p$ is predicted which each phase exhibit distinct topological characteristic. In the next section we examine the influence of other essential degrees of freedom on exotic properties of the (BHT-Ni)$_p$ and (BHT-Ni)$_l$.

## 7. The impact of geometrical symmetry breaking on topological behavior

In the preceding sections we have analyzed the effect of charge and spin degrees of freedom variation on the electronic and topological properties. Here, we focus on the impact of adjusting lattice and orbital DOFs. To this end, by replacing half of the sulfur ligands in (BHT-Ni)$_p$ with selenium, we obtain the cis-like and trans-like configurations, hereafter denoted as cis-(BHT-Ni)$_p$ and trans-(BHT-Ni)$_p$. The entire real-space structures of cis-(BHT-Ni)$_p$ and trans-(BHT-Ni)$_p$ along with related unit cells are depicted in Figures (8a, 8c) and (8b, 8d), respectively. Note that the S and Se ligands are coordinated to the central Ni$^{2+}$ ion in a cis-like and trans-like configurations with broken and preserved SIS, respectively. To assess the stability of the cis-(BHT-Ni)$_p$ and trans-(BHT-Ni)$_p$, we rely on the calculated cohesive energy ($E_C$ as in eq 1), which are reported in Table 2. Compared to graphene, both structures exhibit good stability.

After designing the cis-like and trans-like structures, we consider the effect of adjusting the lattice and orbital DOFs on the electronic structure of reduced symmetry structures in pristine and low electron doping counterparts. In the trans-(BHT-Ni)$_p$ the crystal symmetry reduces to C$_{6h}$ in Schönflies notation, corresponding to the P6/m space group (No.175) which is a subgroup of D$_{6h}$. Some mirror symmetries are broken but the SIS is retained. In contrast, in the cis-(BHT-Ni)$_p$ with the elimination of SIS, the crystal symmetry reduces to D$_{3h}$ in Schönflies notation and corresponds to the P$\bar{6}$2m space group (No. 189).

Remarkably, the analysis of the projected density of states for each element in trans-(BHT-Ni)$_p$ and cis-(BHT-Ni)$_p$ structures, compared to (BHT-Ni)$_p$ clearly shows that the Kagome bands of (BHT-Ni)$_p$ near the Fermi level are dominated by the S p$_z$ orbitals, Ni d$_{xz}$ and d$_{yz}$ orbitals with a minor contribution from C p$_z$ orbitals. However, there is a more significant contribution from Se p$_z$ orbitals in comparison with sulfur atoms near the Fermi level of cis-like and trans-like structures.

See Figures 9a, 9b, and 9c for the projected density of states of (BHT-Ni)$_p$, trans-(BHT-Ni)$_p$, and cis-(BHT-Ni)$_p$, respectively.

Figure 10 shows the band structure for both cis-like and trans-like structures without including SOC. Both cis-(BHT-Ni)$_p$ and trans-(BHT-Ni)$_p$ retaining their semiconductor nature and the Dirac point is preserved in trans-(BHT-Ni)$_p$ (Figure 10a). However, breaking of the SIS in cis-(BHT-Ni)$_p$ leads to a gap opening of 64 meV at the Dirac point (see Figure 10b). Before accounting for SOC effects, the bands are spin degenerate as (BHT-Ni)$_p$, consistent with the diamagnetic nature of the square planar complexes with a d$^8$ electron configuration. Upon including SOC (see Figure 10c and 10d), induced gap openings are observed in trans-(BHT-Ni)$_p$ at the Dirac K point, and Γ point ($\Delta_1$ = 12.6 meV and $\Delta_2$ = 16.3 meV as local gap, respectively) and a global gap at the Γ point ($\Delta_3$ = 12.9 meV) whereas those for cis-(BHT-Ni)$_p$ are opened at the Dirac K point and Γ point ($\Delta_1$ = 52 meV and $\Delta_2$ = 15.9 meV as local gap, respectively) and a global gap at Γ point with ($\Delta_3$ = 12.5 meV).

Two distinct band gaps emerge in the cis-like configuration: one originates from SIS breaking (Figure 10b), and the other arises due to SOC in the presence of SIS breaking (Figure 10d). Apart from gap openings with SOC, by comparing the band gaps of trans-like structure with the (BHT-Ni)$_p$, it become evident that the nontrivial Dirac gap in trans-(BHT-Ni)$_p$ is relatively small (14 meV vs. 12.6 meV). A possible explanation could be the existence of hybridization between the Ni d-orbitals and p-orbitals of light atoms (C, S and Se atoms) that affects the strength of SOC [92]. In addition the analysis of the projected density of states shows that in trans-(BHT-Ni)$_p$ structure apart from Ni d$_{xz}$, d$_{yz}$ and S, C p$_z$ orbitals, there are larger contributions in the energy bands from the pz-orbital components of Se atoms than those for S atoms. Consequently, the larger atomic radius of Se atoms, as compared to S atoms, lead to an increase in hybridization and so a decrease in the SOC Dirac gap. For comparison of SOC gaps of low electron doping counterparts of the (BHT-Ni)$_p$ and trans-(BHT-Ni)$_p$ see Table S2.

In the cis-like configuration with broken SIS and including SOC, Zeeman-type spin splitting of energy bands is observed for all $k$-points in the BZ, except for special points like M and Γ, which is marked as a blue circle in Figure 10d. Although upon including SOC, the spin-up and spin-down bands are no longer separable, the z component of the spin ($\hat{s}_z$) is approximately a conserved quantum number close to the K and K′ points. Therefore, using the projection of spin operator $\hat{s}_z$ —i.e. $\langle\psi_{nk}|\hat{s}_z|\psi_{nk}\rangle$— acquired from Wannier interpolation, we derived $\hat{s}_z$ projected band structure of cis-(BHT-Ni)$_l$ as shown in Figure 11a. In general, under SOC, SIS breaking lifts the spin degeneracy of energy bands at two valleys K and K′. Due to TRS, the spin splitting in opposite valleys must be reversed as shown in Figure 11b, hence, the spin moments can also be used to identify the valley carriers. This forms the foundation of coupled spin and valley physics. In this respect there is a direct band gap at the two inequivalent corners K and K′ of the Brillouin zone. It should be notice that the spin splitting, occur in both valence band maximum and conduction band minimum of cis-(BHT-Ni)$_l$ (see Figure 11b) unlike the spin splitting in MoS$_2$ [93] where, only the valence band maximum is spin-splited. As the conduction band minimum at MoS$_2$ is made of the Mo ($d_z^2$) orbital, the SOC is inactive and the conduction band minimum remains spin degenerate. The magnitude of spin splitting depends on the relative strength of SOC in materials.

Next, we analyze the relevant topological properties and in particular, consider the intrinsic contribution of SHC, which is independent of any scattering. As (BHT-Ni)$_l$ the trans-(BHT-Ni)$_l$ also have large SHC by moving the chemical potential to the SOC Dirac gap (see Figure 12a and 12b). The calculated SHC ($\sigma_{xy}^z$) = -325 ($\hbar/e$)(S/cm) for trans-(BHT-Ni)$_l$ indicates the preservation of quantization (see Figure 12b). In order to determine the SHA, we calculated the longitudinal electrical conductivity $\sigma_{yy}$ using the Boltzmann transport equation. It is important to mention that the shift in $E_F$ results in a change in electron concentration in the calculation of EC. Different elements of EC and the SHA of trans-(BHT-Ni)$_p$ and trans-(BHT-Ni)$_l$, are reported in Table 4. A comparison of the reported data on SHA in Table 3 and Table 4 reveals a higher SHA in trans-(BHT-Ni)$_l$ compared to the (BHT-Ni)$_l$. In addition to the quantized SHC, the edge states connecting the bulk states are another reason for maintaining the QSH phase in the trans-(BHT-Ni)$_l$ (see Figure S4 in the SI file).

Although in the presence of SIS breaking in cis-(BHT-Ni)$_l$ the SHC disappears, another topological characteristic emerges. In cis-(BHT-Ni)$_l$ structure the charge carriers also acquire a valley-contrasting Berry curvature [94]. The BC ($\Omega_{n,xy}^c(\boldsymbol{k})$) of the occupied states can be written as eq 4. According to Vanderbilt [64] if a crystal exhibits both SIS and TRS, then BC is zero. Following this point by breaking SIS in cis-(BHT-Ni)$_l$ the BC is appeared (see Figure 13a) as we have seen regarding TRS breaking in (BHT-Ni)$_h$ (see Figure 7f). The $\Omega_{xy}^c(\boldsymbol{k})$ of cis-(BHT-Ni)$_l$ was significantly peaked at both K and K′ (Figure 13a) but with opposite signs ($\Omega_{xy}^c(K) = - \Omega_{xy}^c(K')$) [94]. The $\boldsymbol{k}$-space contrasting BC in systems without SIS is a key quantity to characterize the chirality of the Bloch electrons and is the basis for valley-contrasting phenomena like VHE [94, 95]. Away from the two valleys, $\Omega_{xy}^c$ decays rapidly and vanishes at the Γ and M points (see Figure 13b). According to eq 4 there exist reverse correlation between the gap value and BC magnitude. So, a tiny band gap (see table S2) in cis-(BHT-Ni)$_l$ has resulted in a large BC with considering SOC ($10 \times 10^3 Å^2$), namely the smaller the SOC gaps the bigger the created BC. Interestingly, the BC magnitude which is shown in Figure 14a are no longer zero in the absence of SOC. The distribution of the BC in 2D momentum space of the valence band maximum is illustrated in Figure 14b. Therefore, a VHE is triggered in planar cis-(BHT-Ni)$_l$ structure even without accounting for SOC. This behavior is different from that in the transition-metal dichalcogenides and similar materials in which SOC is considered as an essential factor to produce the VHE [96]. Breaking SIS leads to two asymmetric sides in the cis-like configuration. The calculated edge states for two asymmetric sides of cis-(BHT-Ni)$_l$ is an indication for the SIS breaking and VHE appearance (see Figure S4 in SI file). The magnitude of BC without including SOC (see Figure 14a), compared to the one with including SOC (see Figure 13a) is smaller. This is due to the fact that, as previously mentioned, when SOC is considered, spin degeneracy is lifted at special points (the K and K′ points in Figure 11a), resulting in smaller band gaps around the Fermi level ($E_F$). So, the peaks of BC with SOC, will be higher than those without SOC.

## 8. Conclusions

The electronic and topological properties of several BHT-Ni materials have been studied and the influence of lattice-, charge-, spin-, orbital- and valley degrees of freedom assessed. Manipulating the charge relocates the Fermi level to the SOC gaps to achieve the desired topological properties. This requires doping with electrons or holes, depending on the topology of the band structure in the material. In the case of (BHT-Ni)$_p$ two (or four) electrons per unit cell are needed to shift the Fermi level inside the SOC gaps, which corresponds to electron doping concentration of up to ∼2 × 10$^{14}$ cm$^{-2}$ [14]. To evaluate this, we investigated the topological aspects of electron doping in different concentrations. The QSH state appeared at low electron doping concentration (1.07×10$^{14}$ cm$^{-2}$) which the calculated helical edge states, quantized SHC, in addition to the nonzero Z$_2$ topological invariant, overall confirm the QSH state in (BHT-Ni)$_l$.

The analysis of the SHC in 2D (BHT-Ni)$_l$, indicates that $\sigma_{xy}^z$ moves to the maximum value of -325 ($\hbar$/e)(S/cm). A large SHA is found in low electron doped counterpart (BHT-Ni)$_l$. Further increasing electron doping concentration, affects the magnetic properties. Thus, with four electrons per unit cell, four spin polarized Kagome bands fill and the Fermi level shifts to the Dirac point of spin down channel.

Breaking TRS (E ($k$, ↑) ≠ E (-$k$, ↓)), leads to a large BC at both K and K′ points with peaks sharing the same sign for spin down channel at Fermi level. But the BC at the spin up channel is obtained with opposite sign. The calculated chiral edge states, quantized AHC, in addition to the integer Chern number, collectively validate the presence of Chern insulating state in (BHT-Ni)$_h$.

Tuning the degrees of freedom allows one to engineer the Berry curvature, as a key topological aspect. Initially, by adding two electrons SBC was created, then by raising the electron doping concentration to four electrons, SBC converted to the BC with the same sign at all BZ corners. This occurs, because of the SIS that implies $\Omega(k) = \Omega(-k)$ and finally by breaking the SIS in cis-(BHT-Ni)$_l$ and alternating the lattice and orbital DOFs, BC changes its sign at K′. The change in the sign of BC is attributed to the presence of TRS that implies $\Omega(k) = -\Omega(-k)$. Furthermore, the spin splitting of bands is another exotic property that appears by adjusting the lattice and orbital DOFs, that is non-uniform throughout the $k$ space. It peaks at the K and K′ points and diminishes away from them. Finally, modifying the lattice and orbital DOFs at trans-(BHT-Ni)$_l$ configuration, leads to higher SHA and preservation of nontrivial topological properties.

Summing up, the already synthesized trivial (BHT-Ni)$_p$ shows transition between different trivial and nontrivial topological states, by altering different DOFs. In this transition, QSH state appears at low electron doping concentration ($1.07 \times 10^{14}$ cm$^{-2}$) and that of QAH state at high ($2.15 \times 10^{14}$ cm$^{-2}$) electron doping levels. At first, QSH state characterized by a nonzero $Z_2$ topological invariant, and then TRS breaking leads to a robust QAH state with a nonzero Chern number. With simultaneous realization of these phases, different types of topological transports corresponding to these phases also observed. The quantized SHC and AHC corresponding to (BHT-Ni)$_l$ and (BHT-Ni)$_h$ respectively, confirm their topologically nontrivial states. Despite small SHC of (BHT-Ni)$_l$, compared to heavy metals, its small electrical conductivity leads to large SHA as other TIs. Our theoretical investigations of (BHT-Ni)$_l$ not only reveal the interaction between the intrinsic SHC and band topology but also offer a promising material foundation for the future implementation of spintronic devices. Table 5 provides a summary of the results from this study.

## AUTHOR INFORMATION

**Corresponding Authors**

**Francesc Illas** - *Departament de Ciència de Materials i Química Física & Institut de Química Teòrica i Computacional (IQTCUB), Universitat de Barcelona, C. Martí i Franquès 1, 08028 Barcelona, Spain.*

**Fariba Nazari** - *Department of Chemistry, Institute for Advanced Studies in Basic Sciences, Zanjan 45137-66731, Iran.*

**Authors**

**Nafiseh Falsafi** - *Department of Chemistry, Institute for Advanced Studies in Basic Sciences, Zanjan 45137-66731, Iran.*

**Saeed H. Abedinpour** - *Department of Physics, Institute for Advanced Studies in Basic Sciences, Zanjan 45137-66731, Iran.*

## COMPETING INTERESTS

There are no conflicts of interest to declare.

## ACKNOWLEDGMENTS

F.N. is grateful to the *Institute for Advanced Studies in Basic Sciences* for financial support through research Grant No. G2023IASBS32604. The research at the *Universitat de Barcelona* has been supported by the Spanish Ministry of Science, Innovation and Universities (MICIUN) Spanish MCIN/AEI/10.13039/501100011033 PID2021-126076NB-I00 and TED2021-129506B-C22 funded partially by FEDER *Una manera de hacer Europa*, and *María de Maeztu* CEX2021-001202-M grants, including funding from European Union and, in part, by the *Generalitat de Catalunya* grant 2021SGR00079.

**ASSOCIATED CONTENT**

**Supporting Information**

The supporting information is available free of charge at (DOI): Additional information includes the convergence of the *k*-mesh grid for SHC calculations, some band structures, data of SHC for various materials, comparison of SOC gaps in pristine and low electron doping concentration counterparts of the structures that are studied in this work, and semi-infinite band structures of trans- and cis-(BHT-Ni)$_l$ are available in the Supporting Information file.

**Table 1.** Conventional components of the SHC tensor by the number of independent components present in all 230 space groups.

| Independent components | 6 | 3 | 2 | 1 |
|---|---|---|---|---|
| Space group | 1-74 | 75-194 | 195-206 | 207-230 |

**Table 2.** Cohesive energy ($E_C$) per atom (eV) of the (BHT-Ni)$_p$, trans-(BHT-Ni)$_p$ and cis-(BHT-Ni)$_p$ compared to graphene.

| Structure | Graphene | (BHT-Ni)$_p$ | trans-(BHT-Ni)$_p$ | cis-(BHT-Ni)$_p$ |
|---|---|---|---|---|
| $E_C$/Atom | -5.65 | -6.39 | -5.95 | -5.94 |

**Table 3.** SHC of (BHT-Ni)$_p$ and (BHT-Ni)$_l$, in the unit of $(\hbar/e)$ (S/cm) at their $E_F$ and longitudinal elements of the electrical conductivity (EC) in the units of (S/cm) at the $E_F$ and the dimensionless SHA ($\theta_{SH}$) for these two structures.

| Structure | $\sigma_{xy}^z$ | $\sigma_{xx}$ | $\sigma_{yy}$ | SHA |
|---|---|---|---|---|
| (BHT-Ni)$_p$ | -0.093 | 46.80 | 46.81 | -0.004 |
| (BHT-Ni)$_l$ | -325 | 343.76 | 343.88 | -1.89 |

**Table 4.** SHC for trans-(BHT-Ni)$_p$ and trans-(BHT-Ni)$_l$, in the units of $(\hbar/e)$(S/cm) at their $E_F$ and the longitudinal elements of the EC in the unit of (S/cm) at $E_F$. The SHA is also reported.

| Structure | $\sigma^z_{xy}$ | $\sigma_{xx}$ | $\sigma_{yy}$ | SHA |
|---|---|---|---|---|
| Trans-(BHT-Ni)$_p$ | -0.254 | 5.84 | 5.85 | -0.09 |
| Trans-(BHT-Ni)$_l$ | -325 | 332.00 | 331.92 | -1.96 |

**Table 5.** This table offers a quick overview of all the findings from this study and illustrates which new phases—whether trivial or non-trivial topological phases—are created by altering each degree of freedom.

| System | Phase | Symmetry | Edge states | Band | Berry curvature |
|---|---|---|---|---|---|
| (BHT-Ni)$_p$ | Trivial insulator | TRS+SIS | Non | 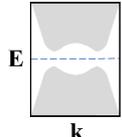 | 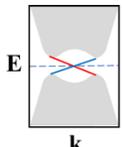 $\Omega(k)=0$ |
| (BHT-Ni)$_l$ | QSH | TRS+SIS | Helical edge states | 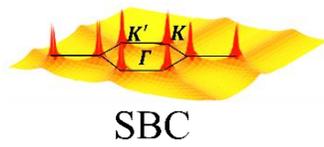 | 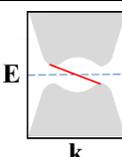 SBC |
| (BHT-Ni)$_h$ | QAH | SIS | Chiral edge states | 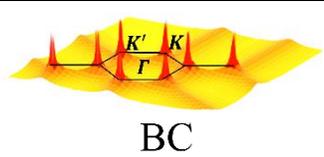 | 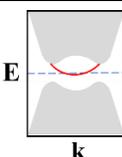 BC |
| Cis-(BHT-Ni)$_l$ | VHE | TRS | Valley polarized edge states | 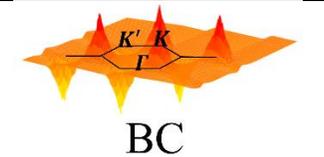 | 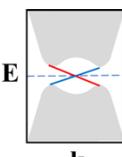 BC |
| Trans-(BHT-Ni)$_l$ | QSH | TRS+SIS+ broken mirror symmetry | Helical edge states | 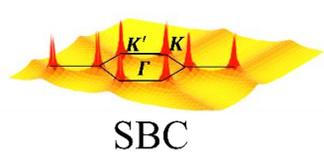 | SBC |

**Figure 1**. (a) Atomic structure of the (BHT-Ni)$_p$ represents the Kagome lattice with black dashed lines. Red solid lines at the middle illustrate the unit cell structure with *a* and *b* lattice vectors. The top left figure shows the motif (Benzene-Hexathiol) that makes up the unit cell. The bottom figure shows the planar optimized structure of 2D MOF (BHT-Ni)$_p$. The C atoms are shown in brown, Ni in gray, and S in yellow. (b) The first BZ and three vertical mirror planes. (c) Band structure without SOC with red Kagome bands and projected density of states of (BHT-Ni)$_p$. The energy at the Fermi level was set to zero.

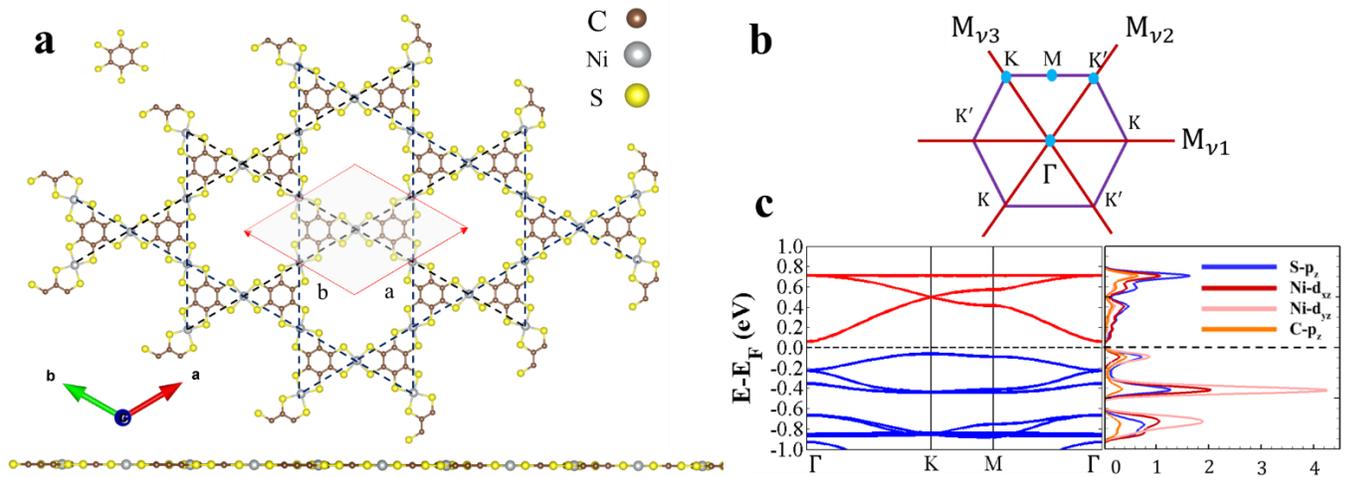

**Figure 2**. (a) The band structure of (BHT-Ni)$_p$ in comparison to (b) (BHT-Ni)$_l$ at the electron doping concentration of $1.07 \times 10^{14}$ cm$^{-2}$, calculated from first principle (DFT) without SOC, (c) The orbital resolved (projected) band structure of the Kagome bands of (BHT-Ni)$_l$ on Ni, S and C atoms respectively and (d) Band structure of (BHT-Ni)$_l$ with SOC and its SOC gaps ($\Delta_1, \Delta_2$ and $\Delta_3$) (e) DFT and MLWF fitted band structures with SOC, obtained from the subspace selected by a projection onto atomic orbitals of (BHT-Ni)$_l$ selected from (c). (f) The semi-infinite helical edge states within the SOC gaps. (g) $k$-resolved SBC on the $k_x$-$k_y$ plane in the BZ at E = E$_F$ by considering SOC.

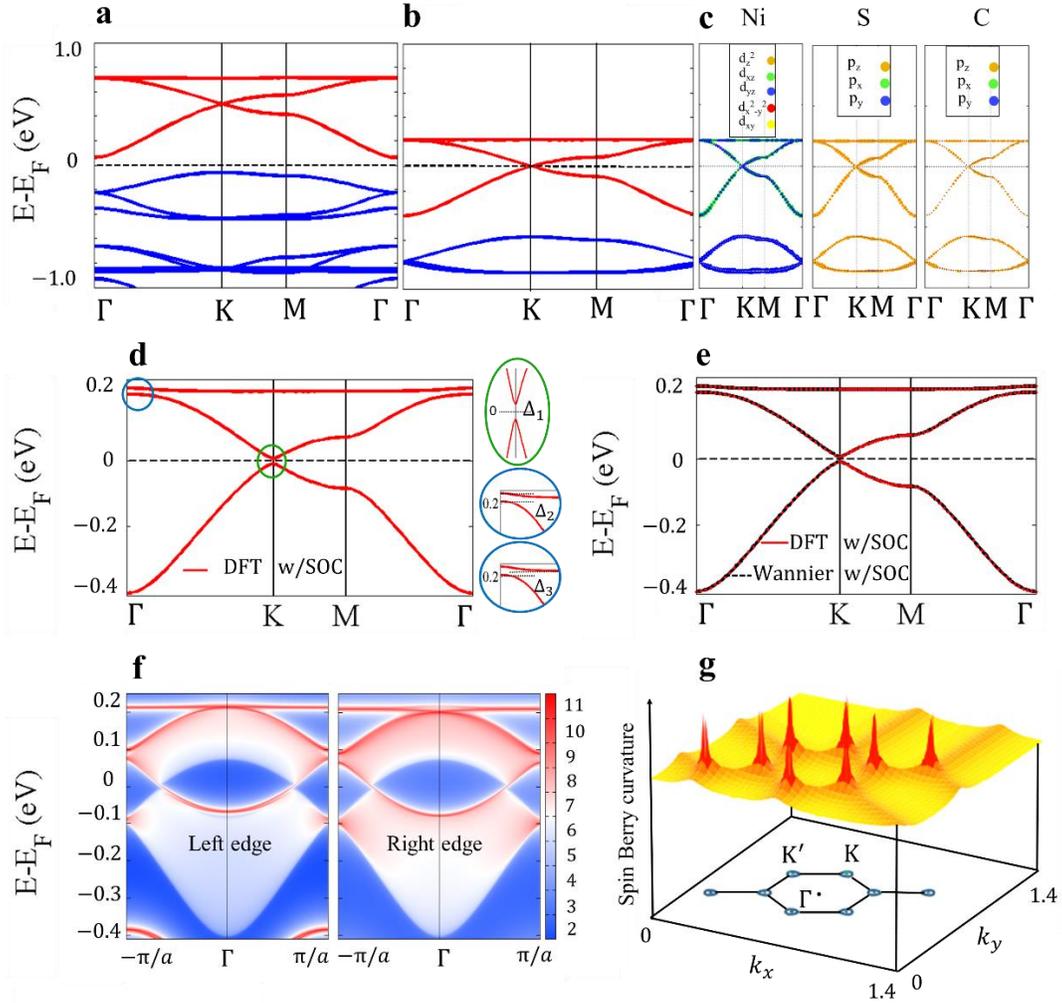

**Figure 3**. (a) band structure of (BHT-Ni)$_p$ in Wannier basis by considering SOC. (b) The $\sigma_{xy}^z$ tensor element of SHC as a function of the E$_F$ position for the (BHT-Ni)$_p$ sheet, that its value in the Fermi level is negligible.

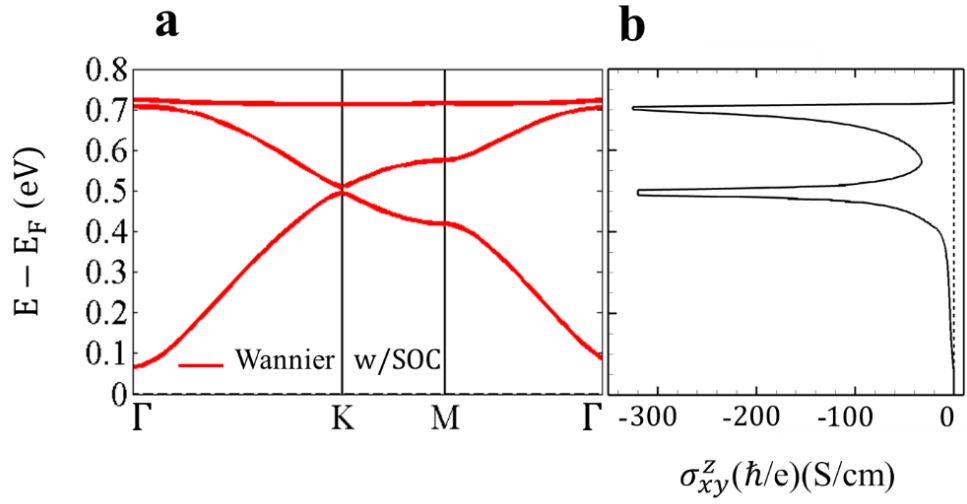

**Figure 4**. (a) band structure of (BHT-Ni)$_l$ in Wannier basis by considering SOC. (b) The $\sigma_{xy}^z$ tensor element of SHC as a function of the E$_F$ position for the (BHT-Ni)$_l$ sheet which illustrates its quantized value at two SOC gaps.

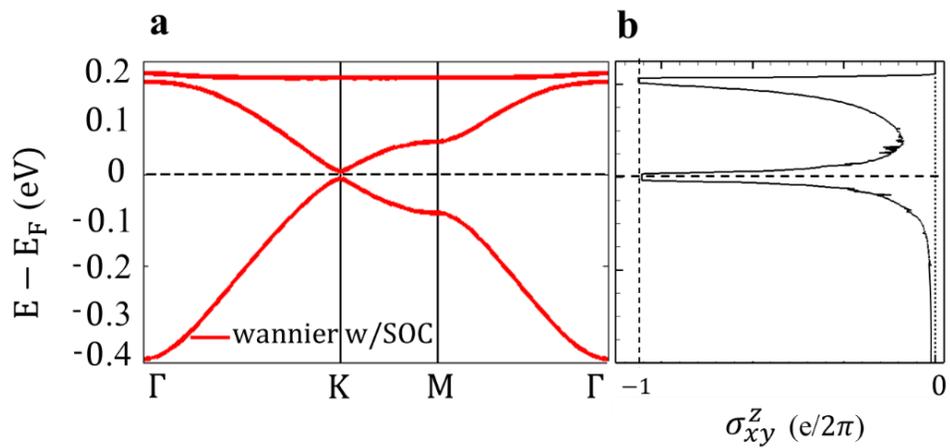

**Figure 5**. The band-projected SBC, $\Omega^z_{n,xy}(\boldsymbol{k})$, in the vicinity of the $E_F$ for (a) (BHT-Ni)$_p$ and (b) (BHT-Ni)$_l$. The energy axis is relative to the Fermi energy denoted by the gray horizontal line. The color of the bands corresponds to the sign and magnitude of the SBC, *i.e.* sgn $(\Omega^z_{n,xy}(\boldsymbol{k}))$ log $|\Omega^z_{n,xy}(\boldsymbol{k})|$.

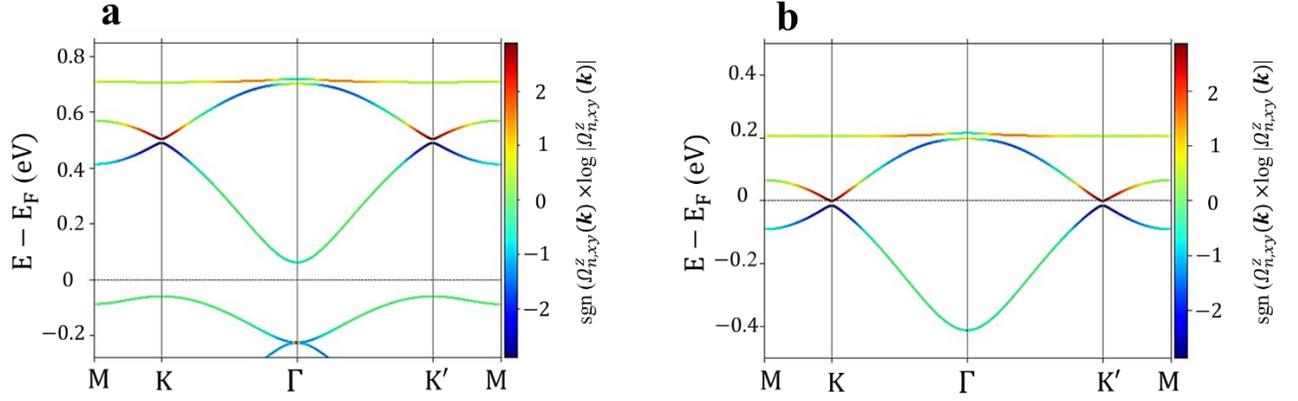

**Figure 6**. The spin-polarized electron density difference (δρ) of (BHT-Ni)$_h$ with an isosurface value of 0.004 Å$^{-3}$.

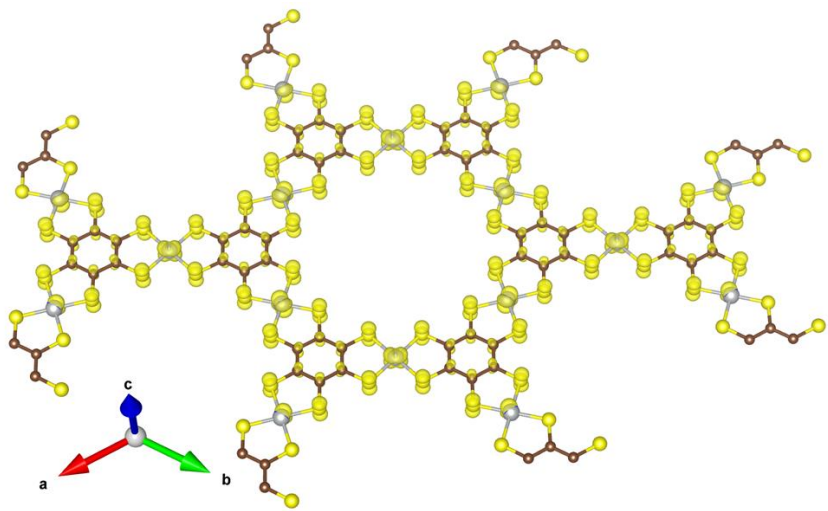

**Figure 7.** (a) The band structure of the (BHT-Ni)$_h$ from first principle (DFT) without SOC. (b) The spin-polarized band structure of the (BHT-Ni)$_h$ calculated without SOC, spin down is red and spin up is blue and (c) Calculation with Wannier basis sets and SOC, six separated Kagome bands and nontrivial gaps with Berry curvatures (red curves) are shown. (d) The calculated quantized AHC and (e) The semi-infinite chiral edge states near the Fermi level based on MLWFs. (f) The distribution of the Berry curvature $\Omega_{xy}^c$ in 2D momentum space for band I (valence band maximum) marked in (c).

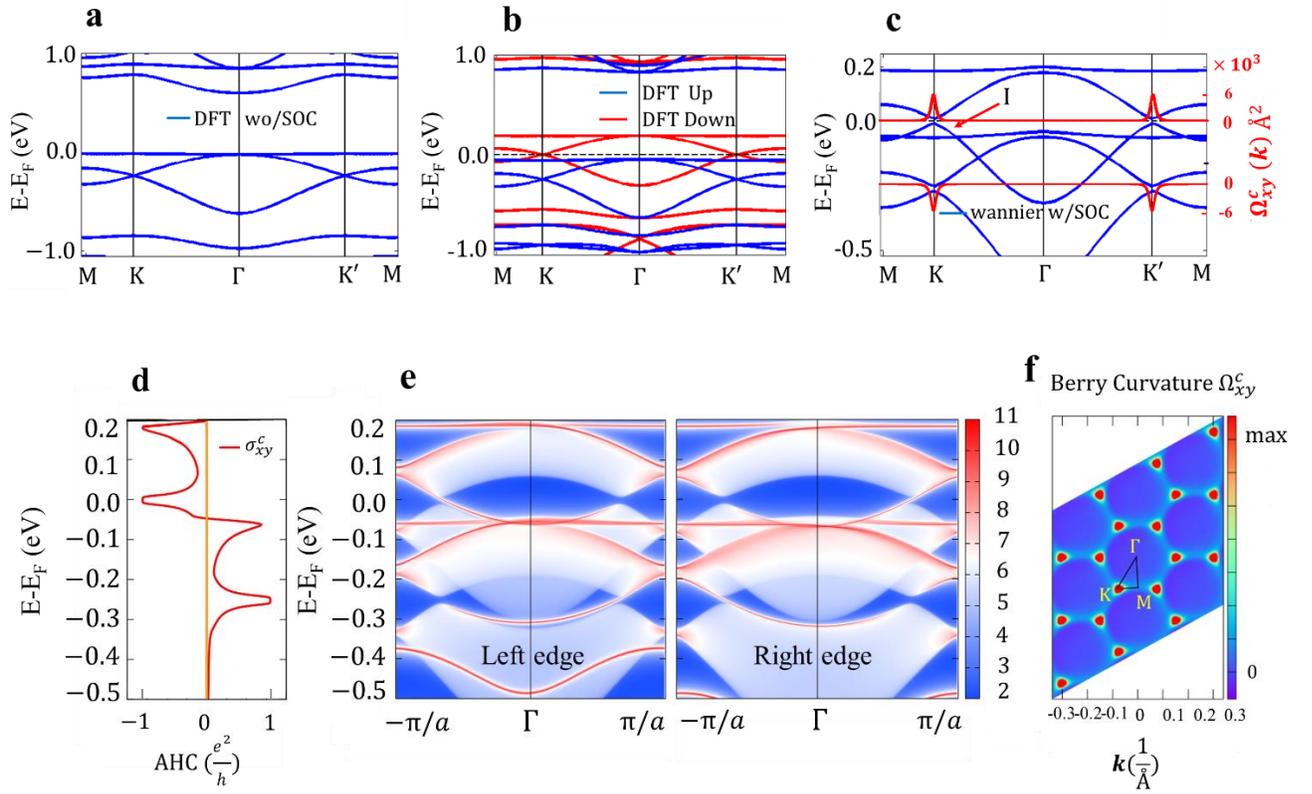

**Figure 8**. Molecular structure of cis- and trans-(BHT-Ni)$_p$ (a) Top view of cis-(BHT-Ni)$_p$ that gray, yellow, red and brown circles are Ni, S, Se and C atoms respectively, the solid line indicates the unit cell. (b) Top view of trans-(BHT-Ni)$_p$, the solid line indicates the unit cell and gray, yellow, red and brown circles are Ni, S, Se and C atoms respectively. (c) Structure of the unit cell (solid line in (**a**)) of the proposed MOFs (M = Ni; X = S, Y = Se) for cis-like configuration and (d) for trans-like configuration.

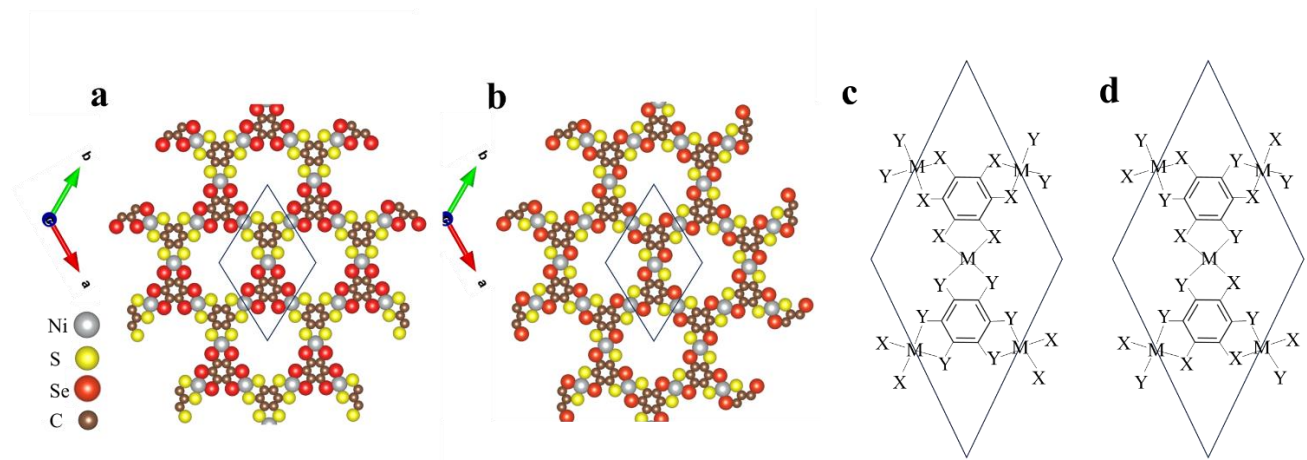

**Figure 9**. Projected density of states of (a) (BHT-Ni)$_p$, (b) trans-(BHT-Ni)$_p$ and (c) cis-(BHT-Ni)$_p$. Kagome bands above the Fermi level are mostly contributed from S p$_z$ orbitals at **a**, Se and S p$_z$ orbitals at **b** and **c**.

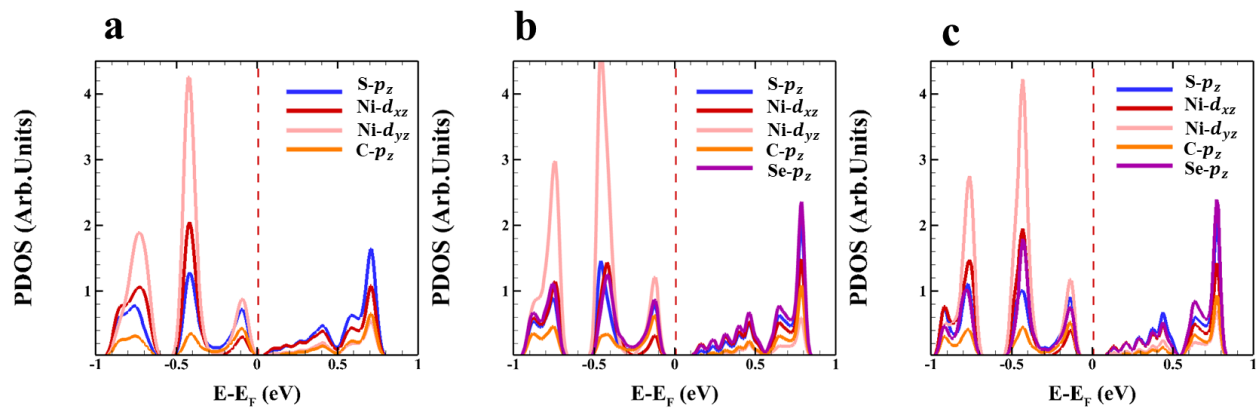

**Figure 10**. Band structure of cis- and trans-(BHT-Ni)$_p$. (a) Band structure of trans- and (b) cis-like configuration without SOC. (c) and (d) Band structures of trans- and cis-(BHT-Ni)$_p$ with SOC. $\Delta_1$ is the Dirac gap, $\Delta_2$ and $\Delta_3$ are local and global gaps at $\Gamma$ point, respectively.

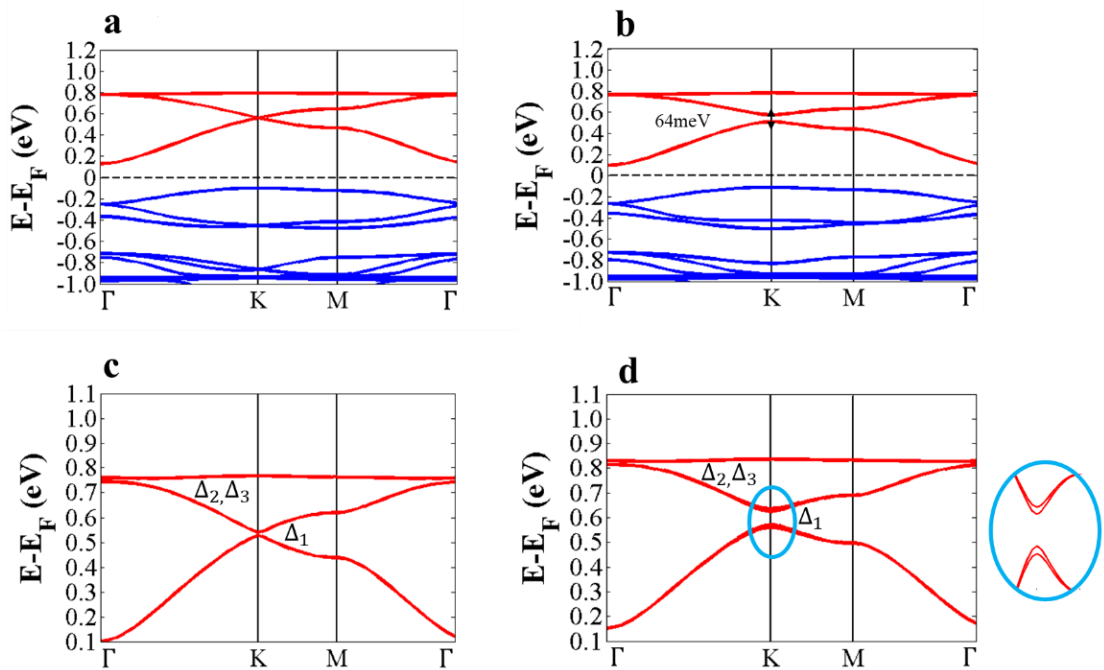

**Figure 11**. (a) The fully relativistic band structure of cis-(BHT-Ni)$_l$ and the projection of spin operator $\hat{s}_z$ (color map). The red and blue colors indicate the spin-up and -down states, respectively. The color scheme is used to show the expectation value $\langle \hat{s}_z \rangle$ of the spin operator $\hat{s}_z$ in the units of $\hbar/2$. (b) Magnification of the parts specified in (**a**).

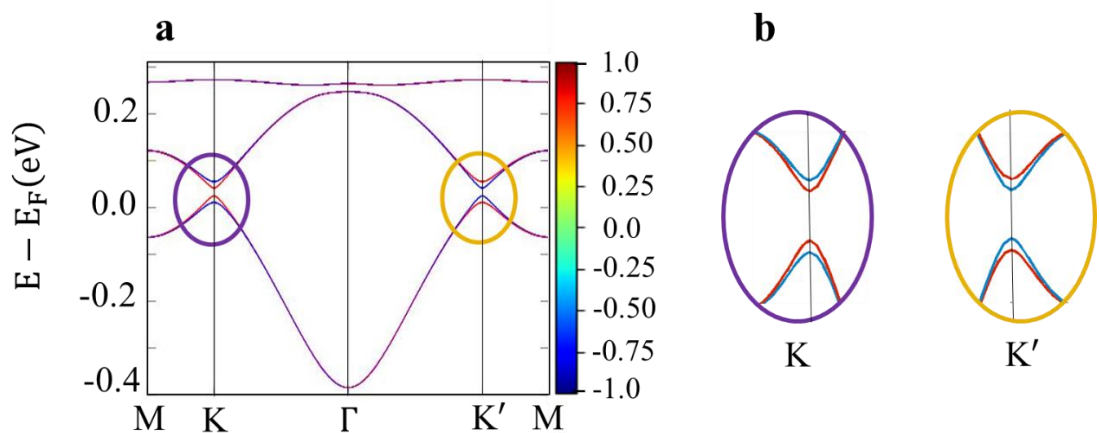

**Figure 12**. (a) Band structure of trans-(BHT-Ni)$_l$ with SOC and (b) The $\sigma_{xy}^z$ tensor element of SHC as a function of the E$_F$ position for the trans-(BHT-Ni)$_l$ sheet, that shows its maximum value at the Fermi level.

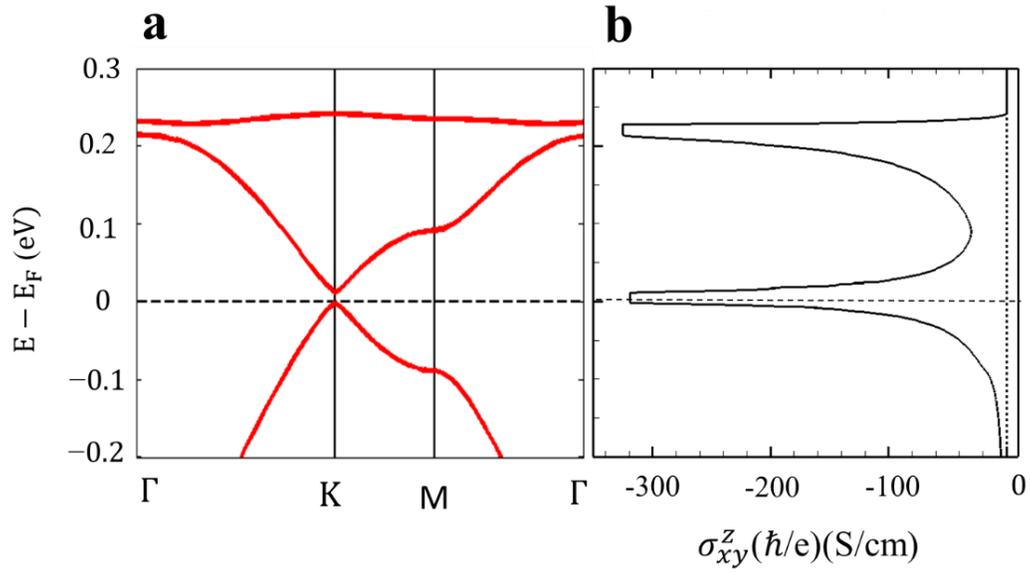

**Figure 13**. (a) The ordinary Berry curvature (BC) of cis-(BHT-Ni)$_l$ with considering SOC, along the high-symmetry lines. (b) The calculated BC distribution of valence bands below Fermi level in the ($k_x$, $k_y$) plane in arbitrary units.

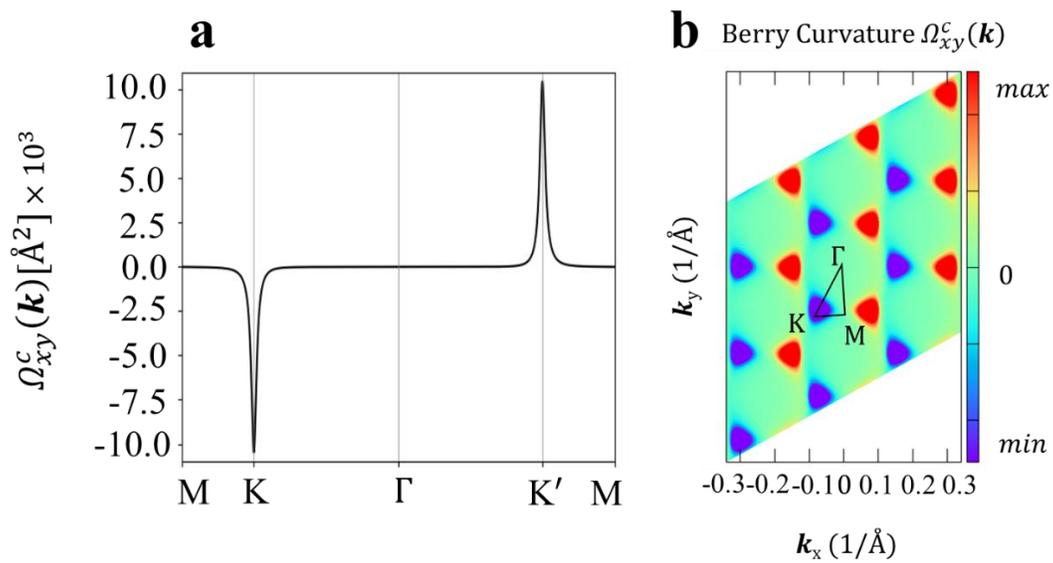

**Figure 14**. (a) The ordinary Berry curvature (BC) of cis-(BHT-Ni)$_l$ without SOC, along the high-symmetry lines. (b) The calculated BC distribution of valence bands below Fermi level in the ($k_x$, $k_y$) plane in arbitrary units.

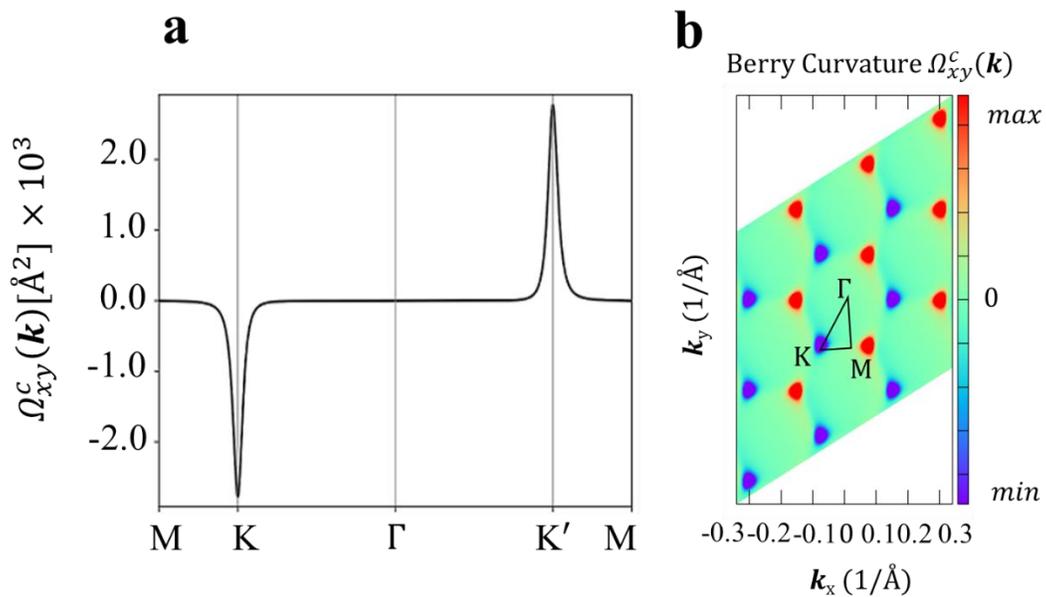

**Graphic for TOC**

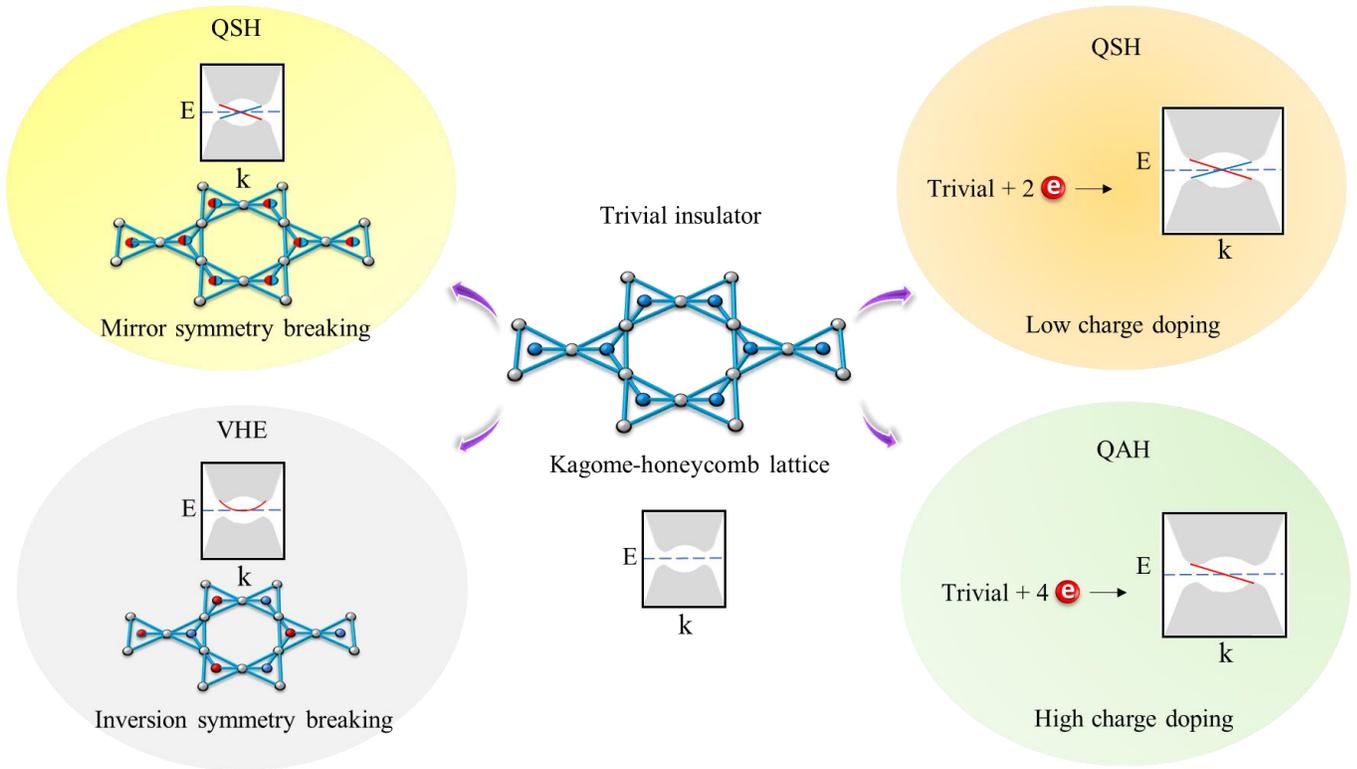